\documentclass[preprintnumbers,amsmath,amssymb,nofootinbib,showkeys,10pt]{revtex4}

\usepackage{graphicx}
\usepackage{subfigure}
\usepackage{dcolumn}
\usepackage{bm}

\begin{document}

\title{Vortex and droplet in holographic D-wave superconductors}

\author{Dongfeng Gao }

\altaffiliation{Email: dfgao@wipm.ac.cn}
\vskip 0.5cm
\affiliation{State Key Laboratory of Magnetic Resonance and Atomic and Molecular Physics, Wuhan Institute of Physics and Mathematics, Chinese Academy of Sciences, Wuhan 430071, China}

\date{\today}

\begin{abstract}
\begin{center}
{\bf ABSTRACT}
\end{center}

We investigate non-trivial localized solutions of the condensate in a (2+1)-dimensional D-wave holographic superconductor model in the presence of a background magnetic field. The calculation is done in the context of the (3+1)-dimensional dual gravity theory of a charged massive spin-2 field in an AdS black hole background. By using numeric techniques, we find both vortex and droplet solutions. These solutions are important for studying the full phase diagram of D-wave superconductors.
\end{abstract}

\maketitle

\section{Introduction}

The holographic principle (also known as the AdS/CFT correspondence) is one of the most significant results from string theory \cite{maldacena1997}, \cite{gubser1998} and \cite{witten1998}. It relates a d-dimensional quantum field theory to a d+1-dimensional gravity theory. One of the exciting aspects of the holographic principle is that it can provide us a powerful tool to study strongly coupled quantum field theories by dealing with their dual classical gravity theories. One application of this tool is to the high temperature superconductor physics.

High temperature superconductors can be classified into two classes. The first class were discovered in 1986 \cite{bednorz1986}. They are cuprates with layered structure and the superconductivity is associated with the copper-oxygen planes. The second class were discovered in 2008 \cite{hosono2008}. They are iron-based superconductors with layered structure as well. The superconductivity is again associated with the two-dimensional planes. However, after more than two decades of experimental and theoretical investigation, the mechanism of high temperature superconductors remains an unsolved mystery \cite{lee2006}. There are experimental evidences showing that electron pairs still form in cuprate superconductors and that the pairing symmetry is a D-wave symmetry \cite{tsuei2000}. Unlike in the case of conventional superconductors, the pairing mechanism here involves strong couplings. Therefore, people hope that the holographic correspondence can give us some insights in high temperature superconductors.

The first holographic model for superconductors was constructed in \cite{hartnoll2008a}. It is an S-wave model because the order parameter is represented by the condensate of a complex scalar field. The temperature is introduced by the Hawking temperature of an AdS black hole.  Below some critical temperature, the condensate develops. Above it, the condensate vanishes. Thus, the superconducting-to-normal phase transition is produced. Following \cite{hartnoll2008a}, various aspects of S-wave models have been extensively studied (see \cite{hartnoll2008b}, \cite{hartnoll2009}, \cite{herzog2009} and  \cite{horowitz2010} for reviews). The construction of S-wave models in string theory and M-theory was given in \cite{gubser2009} and \cite{gauntlett2009}. The behavior of the S-wave superconducting condensate in a magnetic field background was investigated in \cite{albash2008}, \cite{albash2009a}, \cite{albash2009b}, \cite{montull2009}, \cite{maeda2009} and \cite{ge2010}. In particular, the vortex solution was studied in \cite{albash2009a}, \cite{albash2009b}, \cite{montull2009} and \cite{tallarita2010}, and the droplet solution was discussed in \cite{albash2009a} and \cite{albash2009b}. Motivated by studies of S-wave models, holographic models of P-wave superconductors were investigated in \cite{gubser2008}, \cite{roberts2008}, \cite{basu2010}, \cite{ammon2010a} and \cite{ammon2010b}, where an SU(2) Yang-Mills field is coupled to an AdS black hole, and the order parameter is represented by a vector field.

Recently, holographic models for D-wave superconductors were constructed in \cite{chen2010}, \cite{benini2010a} and \cite{benini2010b}. In these models, a charged massive spin-2 field was put in an AdS black hole background. The condensate of this spin-2 field signals the D-wave superconducting phase transition. Some properties of these models were studied in \cite{zeng2010a}, \cite{zeng2010b} and \cite{chen2011}. In this work, we will study the non-trivial localized solutions of the condensate in the presence of a background magnetic field, based on the action given in  \cite{benini2010a} and \cite{benini2010b}. Especially, we are interested in finding the vortex and droplet solutions.

The organization of our paper is as follows. In section II, we will first introduce the model in \cite{benini2010a} and \cite{benini2010b}. For convenience, the AdS black hole metric is written in terms of polar coordinates. Then, equations of motion for the spin-2 field and the gauge field are derived. In section III, we obtain the vortex and droplet solutions by using numeric techniques. Properties of these solutions are discussed as well. In the final section, we conclude with some remarks.

\section{The D-wave holographic superconductor model}

The full 3+1-dimensional gravity theory which is dual to a 2+1-dimensional D-wave superconductor contains the gravity sector and the charged massive spin-2 field sector
\begin{equation}
S=\frac{1}{2 \kappa^2} \int d^4 x \sqrt{-g} \,\, \{ (\mathcal{R}-2 \Lambda)+\mathcal{L}_m\},
\end{equation}
where $\mathcal{R}$ is the Ricci scalar and $\Lambda=-3/L^2$ is the negative cosmological constant with the AdS radius $L$. $\kappa^2=8 \pi G_4$ is the gravitational coupling. $\mathcal{L}_m$ is the Lagrangian for the charged massive spin-2 field and the gauge field. It is well known that writing down a consistent $\mathcal{L}_m$ in a curved spacetime is a very difficult problem. The authors in \cite{benini2010a, benini2010b} have constructed such a Lagrangian in an AdS spacetime, which will be used in this paper
\begin{equation}
\begin{split}
\label{lag1}
\mathcal{L}_m &= - |D_\rho \varphi_{\mu\nu}|^2 + 2|D_\mu \varphi^{\mu\nu}|^2 + |D_\mu \varphi|^2 - \big[ D_\mu \varphi^{*\mu\nu} D_\nu \varphi + \text{c.c.} \big] - m^2 \big( |\varphi_{\mu\nu}|^2 - |\varphi|^2 \big) \\
&\quad +2  R_{\mu\nu\rho\lambda} \varphi^{*\mu\rho} \varphi^{\nu\lambda}
- \frac{1}{d+1} \mathcal{R} | \varphi |^2
- i  q F_{\mu\nu} \varphi^{*\mu\lambda} \varphi^\nu_\lambda - \frac{1}{4} F_{\mu\nu} F^{\mu\nu} \;,
\end{split}
\end{equation}
where $\varphi_{\mu\nu}$ represents the spin-2 field, which is symmetric. We introduce the notation  $\varphi \equiv\varphi_{\mu}^{\mu}$. $D_\mu = \nabla_\mu - i q A_\mu$ is the covariant derivative, which reduces to the familiar $D_\mu = \partial_\mu - i q A_\mu$ in flat spacetimes. $d$ is the bulk spatial dimension, which is 3 in our case. $F_{\mu\nu}$ is the U(1) gauge field strength tensor. As discussed in \cite{benini2010b}, this Lagrangian works well only when the background spacetime is an Einstein manifold, which means
\begin{equation}
\label{mfd}
R_{\mu\nu} = \frac{2 \Lambda}{d-1} \, g_{\mu\nu}  \;.
\end{equation}
This condition restricts us to do calculations in the probe limit \cite{hartnoll2008a}, where the background metric is not perturbed by the spin-2 field and the gauge field.

It is easy to write down the 3+1-dimensional AdS-Schwarzschild planar black hole solution to (\ref{mfd})
\begin{equation}
ds^2 = - \frac{L^2 \alpha^2}{z^2} f(z)\,dt^2 + \frac{L^2}{z^2} (dr^2 + r^2 d\phi^2)+ \frac{L^2}{z^2 f(z)}dz^2,
\end{equation}
where
\begin{equation}
f(z) = 1 - z^3 \,.
\end{equation}
The black hole horizon is at $z=1$, and the boundary is at $z=0$. They are 2-dimensional planes. We have introduced the polar coordinate system ($r$, $\phi$) on the planes. In our notations, $r$ and $z$ are now dimensionless coordinates. The Hawking temperature of this black hole is
\begin{equation}
T = \frac{3 \alpha}{4 \pi} \, .
\label{temp}
\end{equation}

\section{Localized solutions of the D-wave condensate}

According to the AdS/CFT correspondence, the bulk massive spin-2 field $\varphi_{ij}$ is mapped to a spin-2 operator $O_{ij}$ in the boundary conformal field theory. The conformal dimension $\Delta_{ij}$ of $O_{ij}$ is determined by
\begin{equation}
\label{dim}
m^2 L^2=\Delta_{ij}(\Delta_{ij}-3)\, .
\end{equation}

The following ansatz is used in our work. For the gauge field\footnote{Strictly speaking, there is no dynamical gauge field in the dual boundary theory. But as discussed in many papers, e.g. \cite{hartnoll2009}, this problem does not prevent us from introducing spatially dependent magnetic field.}
\begin{equation}
A= A_{\mu}dx^{\mu}=A_t(r,z)\, dt + A_{\phi}(r,z)\, d\phi ,
\end{equation}
and for the spin-2 field,
\begin{equation}
\varphi_{\mu\nu}=\left({\begin{array}{cccc}0&0&0&0 \\ 0&\frac{L^2 \alpha^2}{z^2}\psi(r,z)&i r \frac{L^2 \alpha^2}{z^2}\psi(r,z)& 0 \\ 0&i r \frac{L^2 \alpha^2}{z^2}\psi(r,z)& -r^2 \frac{L^2 \alpha^2}{z^2}\psi(r,z)&0 \\ 0&0&0&0 \\
\end{array}}\right) e^{i (n+2)\phi} .
\end{equation}
Obviously, the ansatz for $\varphi_{\mu\nu}$ satisfies $\varphi=0$. We also require that $D^{\mu}\varphi_{\mu\nu}=0$ in later calculation. The equations of motion for $A_t$, $A_{\phi}$ and $\psi$ are found to be
\begin{align}
\label{eomat}
0&=-8 q^2 L^2 r^2 \psi^2 A_t +r z^2 (r \, \partial_z^2 A_t +\partial_r A_t +r \partial_r^2 A_t), \\
\label{eomaphi}
0&=4 q^2 L^2 r \psi^2 A_{\phi}-z^2 (r f'\, \partial_z A_{\phi}+r f \, \partial_z^2 A_{\phi}-\partial_r A_{\phi} + r \partial_r^2 A_{\phi}),\\
\nonumber 0&=q^2 r^2 z^2 A_t^2 \psi + \alpha^2 f \, \big(-\psi \, (L^2 m^2 r^2 + n^2 z^2 + q^2 z^2 A_{\phi}^2 + q r z^2 \partial_r A_{\phi})\\
\label{eompsi}
&\,\,\,\,\,\, + r z \big(f \, (-2 r \partial_z \psi + r z \partial_z^2 \psi)+z (r f'\, \partial_z\psi + \partial_r\psi + r \partial_r^2\psi)\big)\big).
\end{align}

The behavior of $A_t$, $A_{\phi}$ and $\psi$, near the AdS boundary at $z=0$, contains important information. The series expansions around $z=0$ take the following form:
\begin{align}
A_t(r,z)&=\mu - \rho \, z + A_{t2}(r) z^2 + A_{t3}(r) z^3+ \cdots , \\
A_{\phi}(r,z)&= a_{\phi}(r) + J_{\phi}(r) z + A_{\phi 2}(r) z^2+ A_{\phi 3}(r) z^3 + \cdots,\\
\psi(r,z)&=z^{3-\Delta} \, (\psi^{(s)}(r)+\mathcal{O}(z))+ z^{\Delta} \, \big(\frac{<O(r)>}{2\Delta-3}+\mathcal{O}(z)\big).
\end{align}
From the expansion of $A_t$, we can read out the chemical potential $\mu$ and the related charge density $\rho$. Similarly, from the first two terms in the expansion of $A_{\phi}$, we read out the boundary magnetic field $B=\frac{1}{r}F_{r\phi}|_{z=0}=\partial_r a_{\phi}/r$ and the azimuthal current density $J_{\phi}$. In the expansion of $\psi$, $\psi^{(s)}$ is identified as the source and $<O(r)>$ is identified as the D-wave condensate, whose conformal dimension $\Delta$ is determined by (\ref{dim}). Only one of them can exist at a time, so we focus on $<O(r)>$ by setting $\psi^{(s)}=0$.

Since we work in the probe limit, we can fix $q=1$ without losing generality. For convenience, we also set the AdS radius $L=1$. Then, the free parameters in our model are the temperature $T$ (through the relation (\ref{temp})), the charge density $\rho$ and the mass $m$. Finding solutions to (\ref{eomat}), (\ref{eomaphi}) and (\ref{eompsi}) is highly non-trivial. We have to resort to numerical techniques. To solve the equations numerically, the boundary conditions have to be specified. At the AdS boundary ($z=0$), we impose the conditions $\psi|_{z=0}=0$, $\partial_z A_t|_{z=0}=-\rho$, and $A_{\phi}=r^2 B/2$. At the black hole horizon ($z=1$), we require that $\psi$ and $A_{\phi}$ are finite and that $A_t=0$. We also have to consider how to determine the critical temperature $T_c$. The condensate only exists below $T_c$. Above it, the superconducting state changes to the normal state
\begin{equation}
\psi=0, \,\,\,\, A_t=\mu-\rho z, \,\,\,\, A_{\phi}= \frac{1}{2} r^2 B .
\end{equation}
Therefore, by observing when a non-zero condensate begins to form, we can determine the value of $T_c$ in terms of $\rho$ and $m$. In the following, the Maple 15 package \cite{maple15} is used to numerically solve the system of the above three partial differential equations.

Two types of non-trivial localized solutions of the condensate will be discussed, characterized by their behavior at $r\rightarrow \infty$. The first type is called the vortex solution, where  $<O(r)>\rightarrow$ a non-zero constant at $r\rightarrow \infty$. The other one is called the droplet solution, where $<O(r)>\rightarrow 0$ at $r\rightarrow \infty$.

\subsection{The vortex solutions}

\begin{figure}[h]
\begin{center}
\subfigure[~~The condensate]{\includegraphics[width=0.8\textwidth]{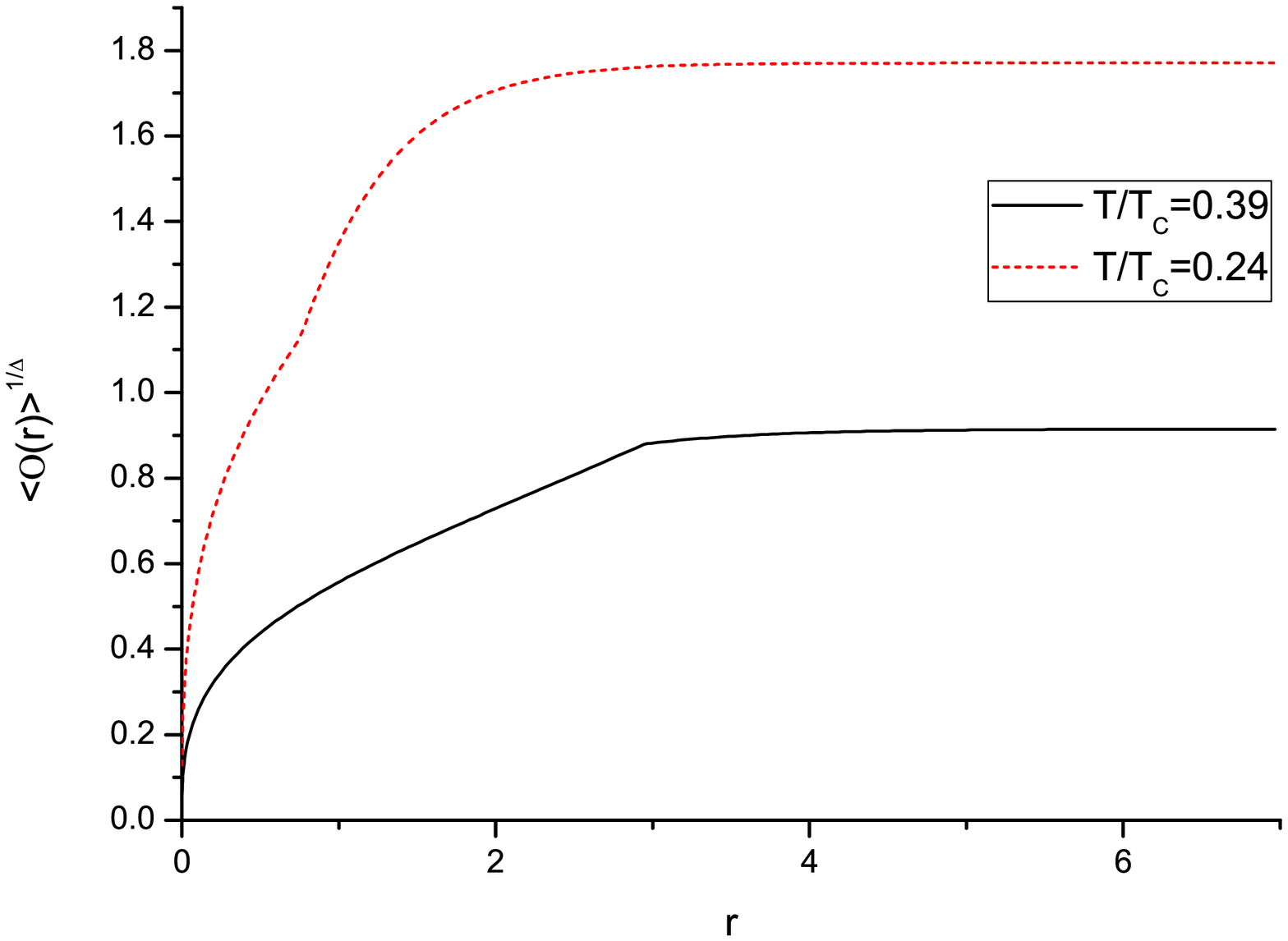}}
\subfigure[~~The magnetic field]{\includegraphics[width=0.8\textwidth]{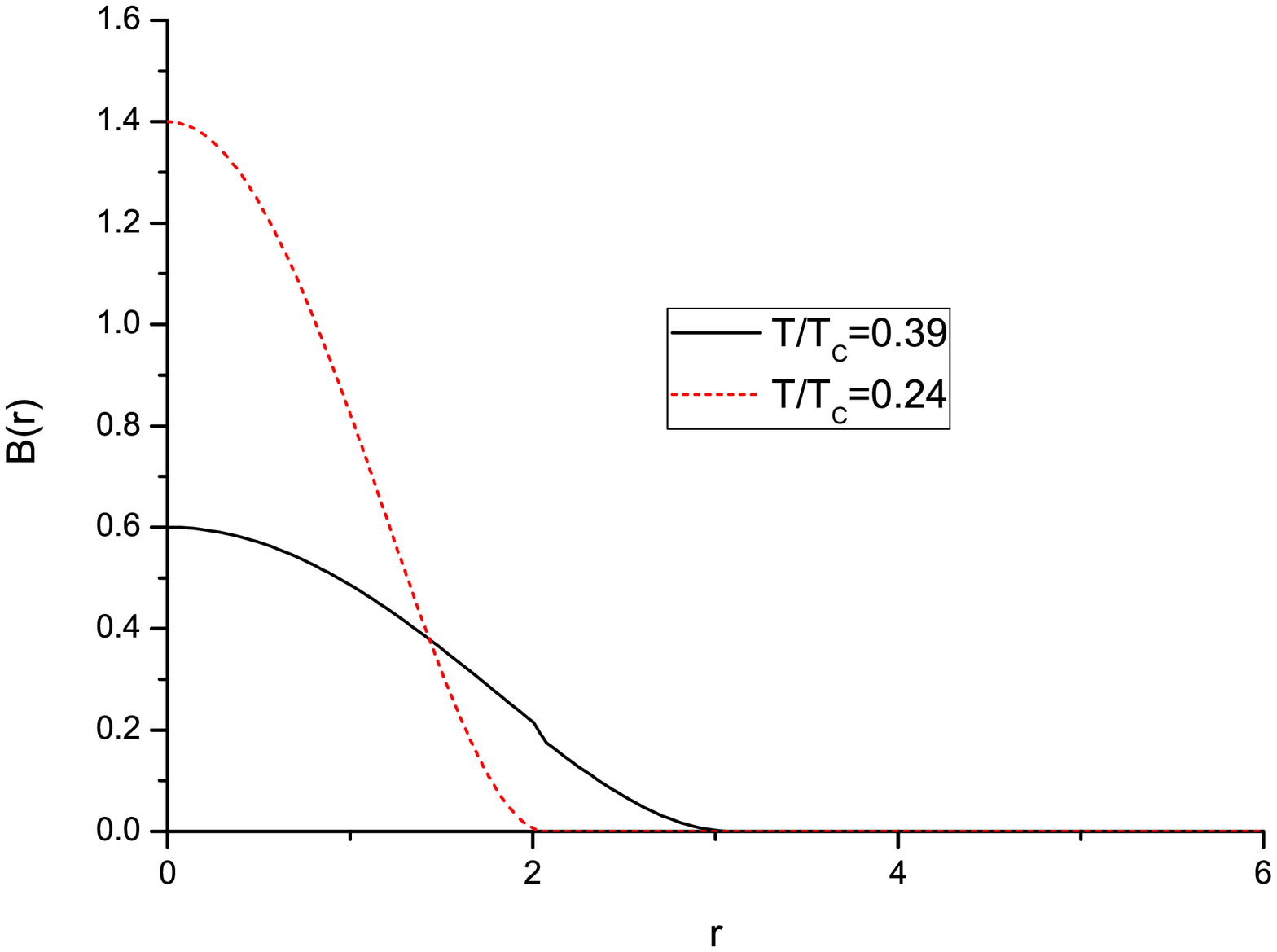}}
\caption{The vortex solution for m=0.025.}
\label{fig1}
\end{center}
\end{figure}

\begin{figure}[h]
\begin{center}
\subfigure[~~The condensate]{\includegraphics[width=0.8\textwidth]{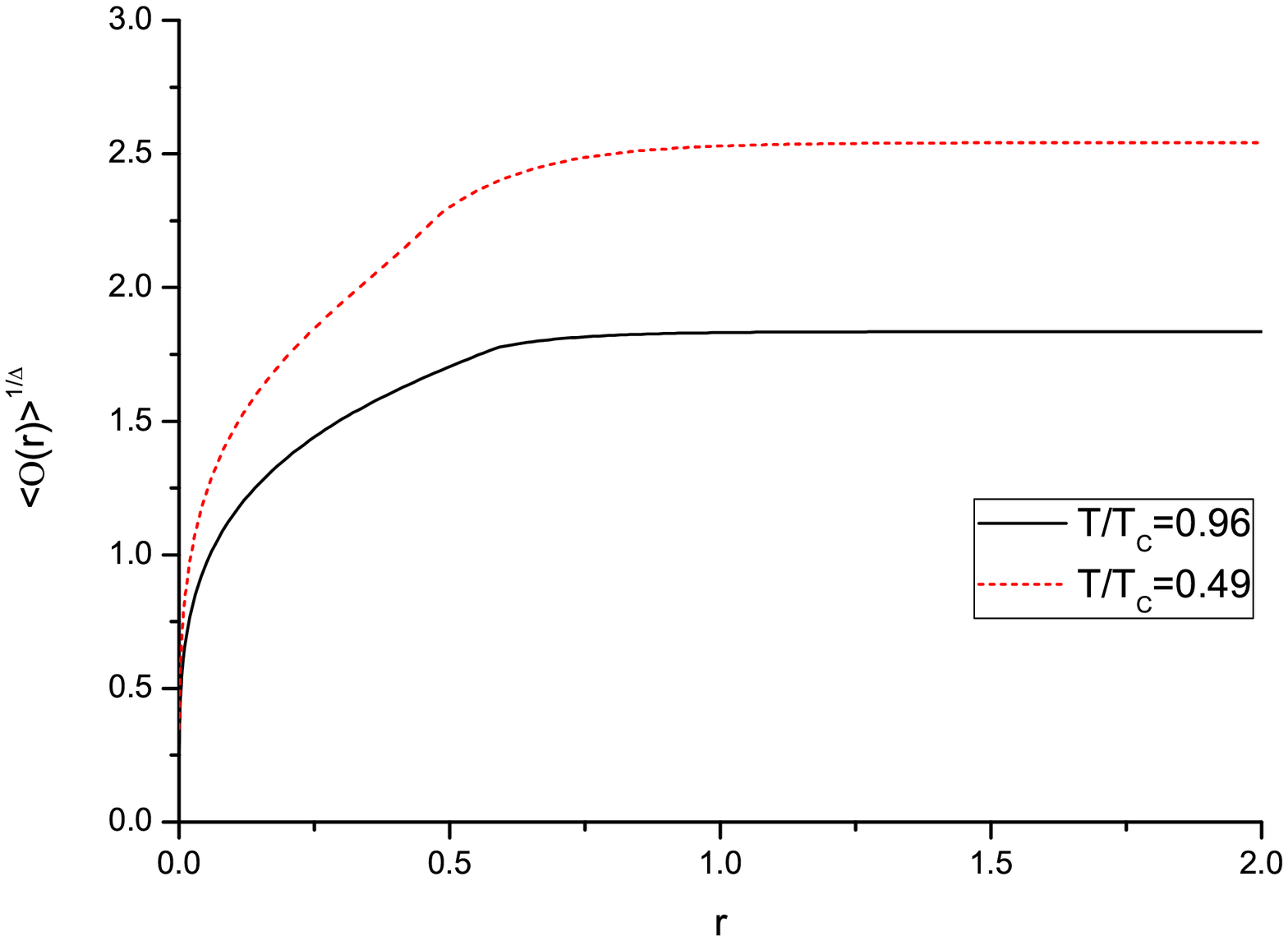}}
\subfigure[~~The magnetic field]{\includegraphics[width=0.8\textwidth]{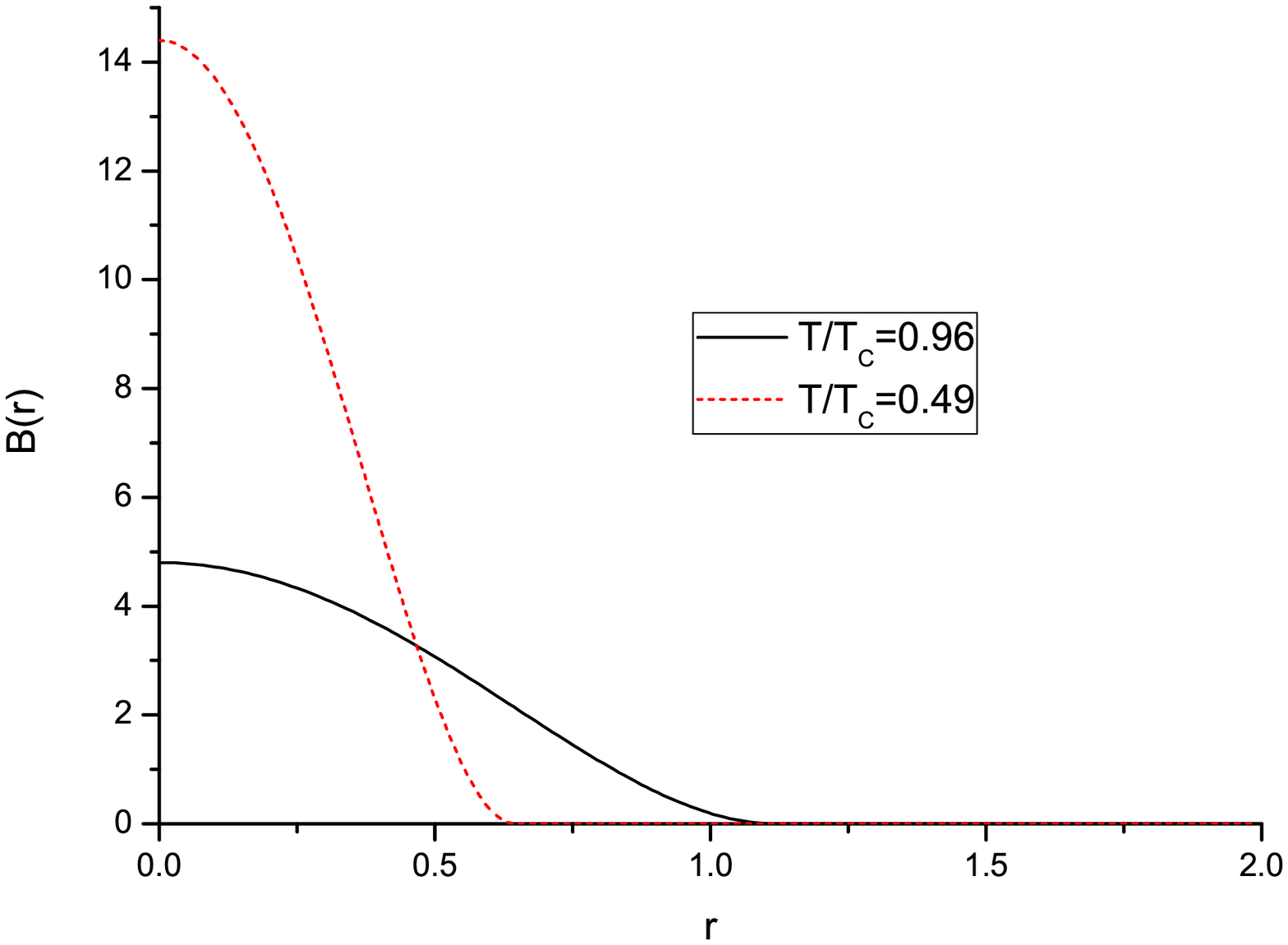}}
\caption{The vortex solution for m=2.0.}
\label{fig2}
\end{center}
\end{figure}

In FIG. \ref{fig1} and \ref{fig2}, we display the vorticity $n=1$ sample solutions for cases of $m=0.025$ and $m=2.0$, respectively. For brevity, only figures for the condensate and the magnetic field are shown in each case. In each figure, for comparison, curves for two different temperatures are drawn. Our result is similar to that of the S-wave holographic superconductors in \cite{albash2009a}, \cite{albash2009b} and \cite{montull2009}. We can see that the condensate goes to zero at the origin where the core of the vortex is located, and runs to a constant quickly when approaching to the infinity. Therefore, the U(1) gauge symmetry is broken at large $r$ as expected for vortex solutions. The magnetic field $B$ starts from a non-zero value at the core of the vortex, and drops down to zero at large $r$. It is obvious that the magnetic field penetrates the superconductor only through the small region around the core of the vortex. The magnetic flux, $\Phi=\int d\phi\int r B dr$, is calculated to be $2\pi$. In both cases, the solutions behave more like vortices at lower temperatures than at higher temperatures.

As pointed out in \cite{benini2010b}, the unitarity requires that $m^2 \geq 0$ in the model. We take two typical values of $m$ for comparison. We find that the vortex solution begins to form at lower temperature for smaller masses than that for larger masses. The condensate grows more quickly for $m=2.0$ than that for $m=0.025$. In other words, It behaves more like a vortex line. This may be due to the reason that the model works better at larger values of $\Delta$. On the other hand, when $\Delta$ becomes large, it becomes difficult to obtain numerical solutions because of the higher order corrections to the action. So, in practice, we should choose moderate values for $m$.

\subsection{The droplet solutions}

\begin{figure}[h]
\begin{center}
\subfigure[~~The condensate]{\includegraphics[width=0.8\textwidth]{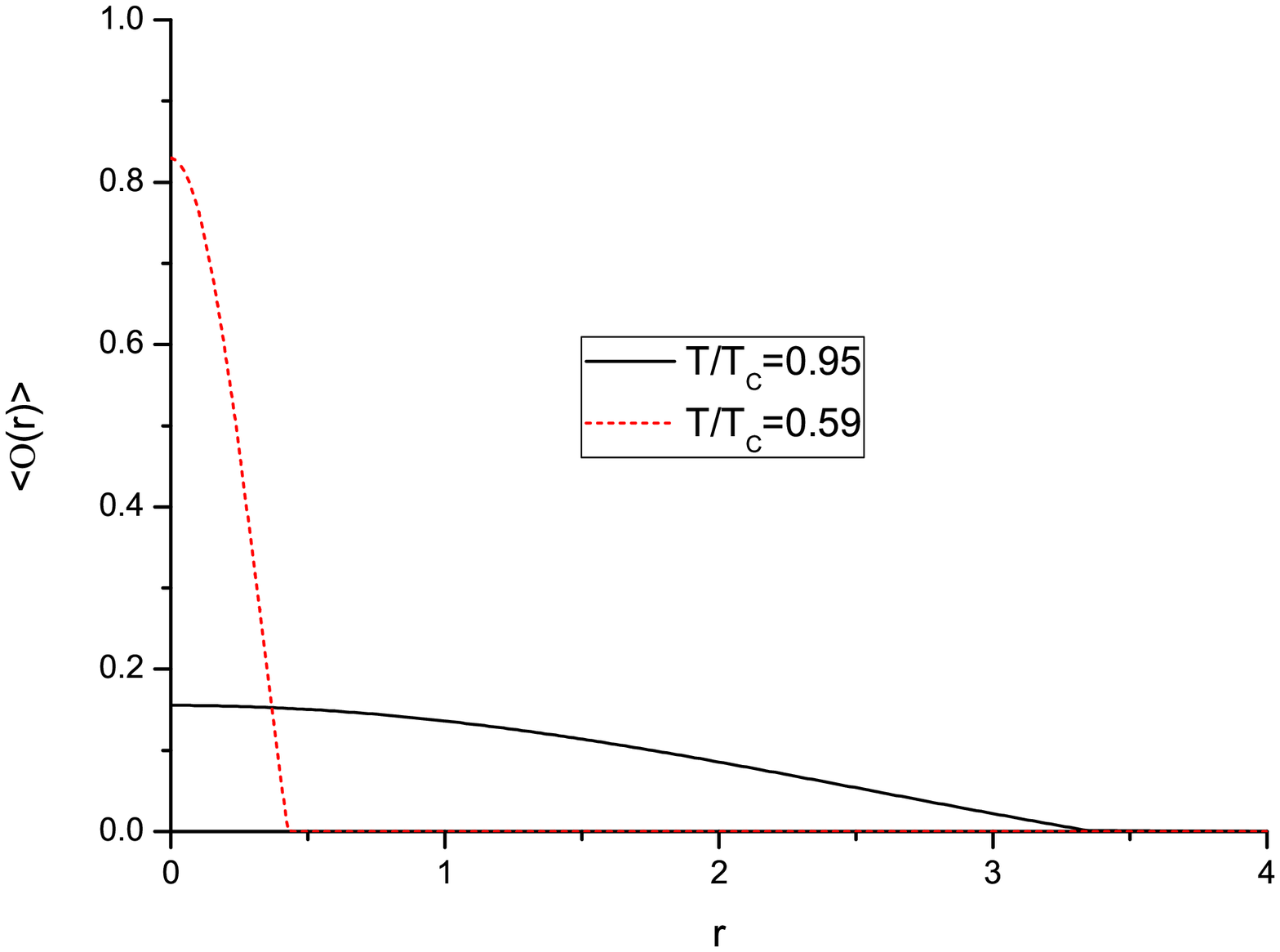}}
\subfigure[~~The magnetic field]{\includegraphics[width=0.8\textwidth]{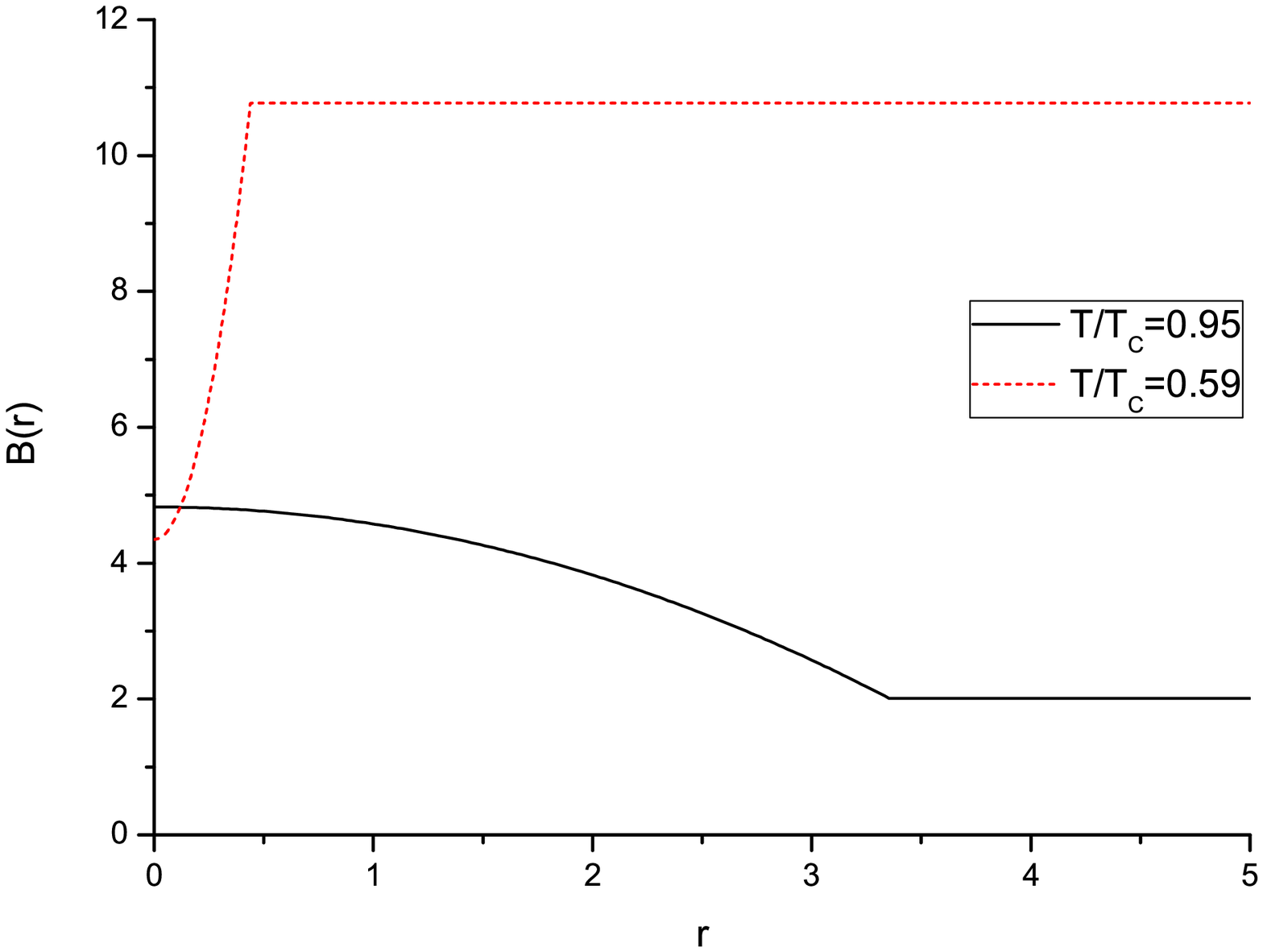}}
\caption{The droplet solution for m=0.025.}
\label{fig3}
\end{center}
\end{figure}

\begin{figure}[h]
\begin{center}
\subfigure[~~The condensate]{\includegraphics[width=0.8\textwidth]{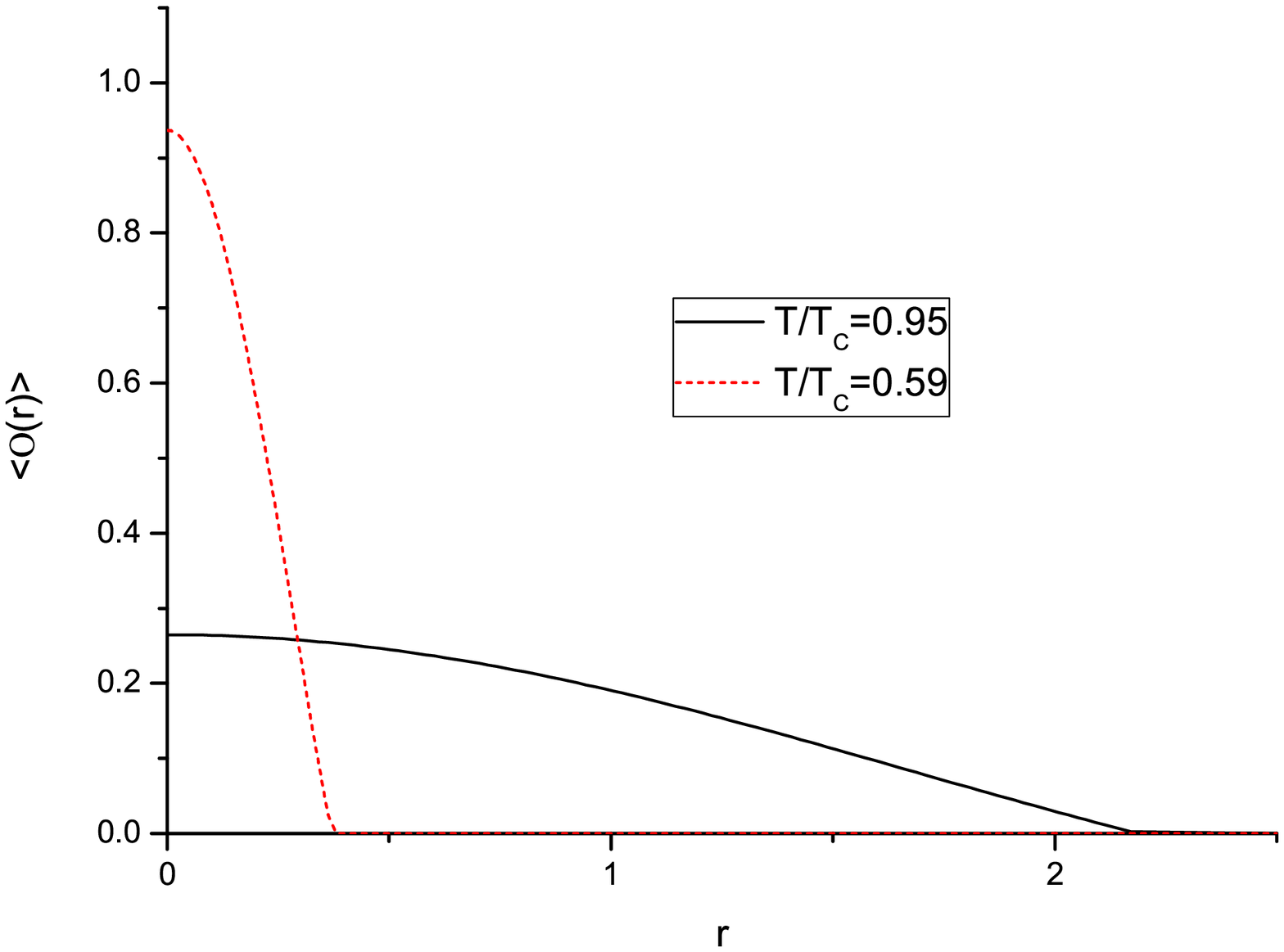}}
\subfigure[~~The magnetic field]{\includegraphics[width=0.8\textwidth]{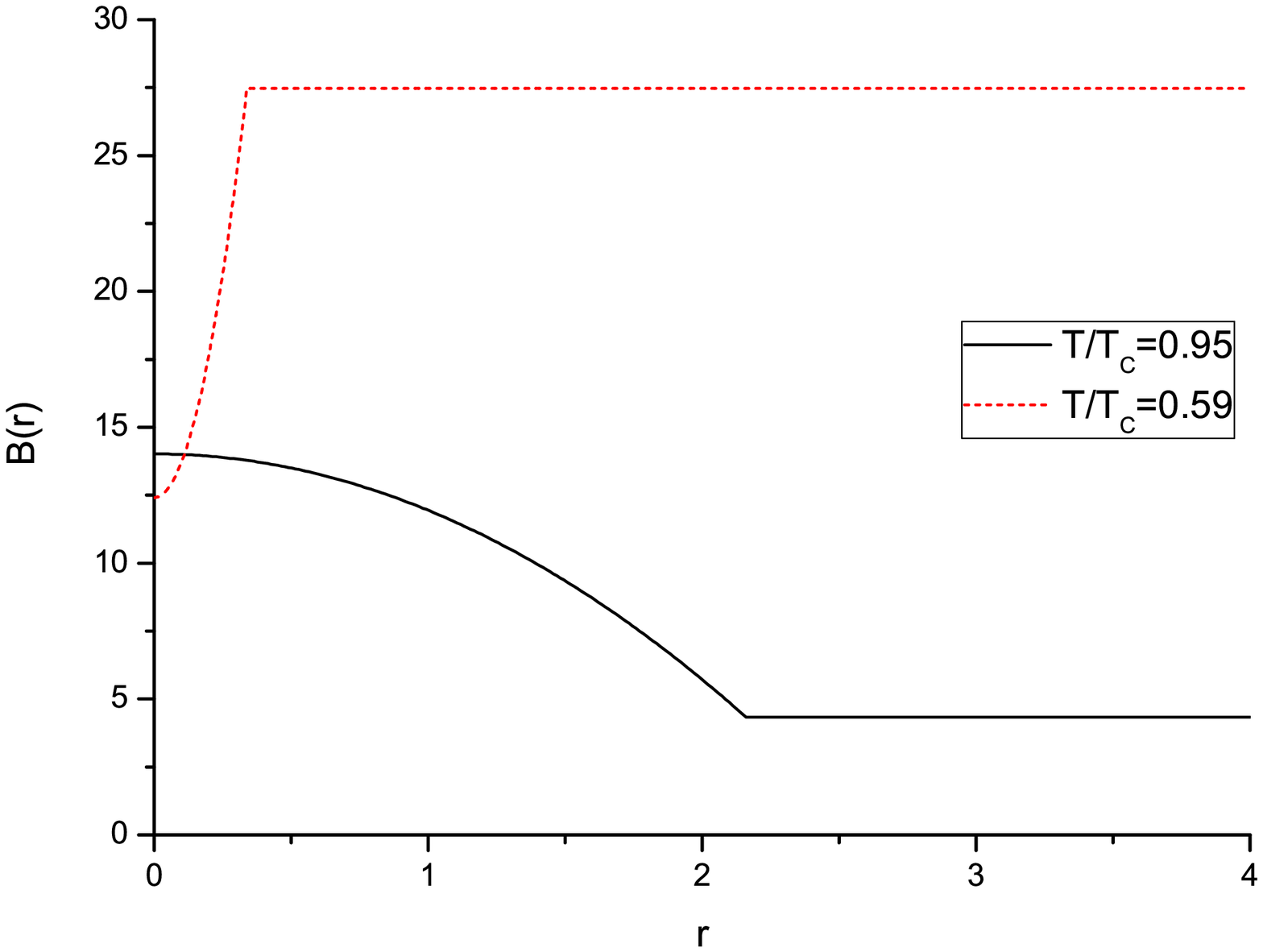}}
\caption{The droplet solution for m=2.0.}
\label{fig4}
\end{center}
\end{figure}

Now, let us discuss the droplet solutions. In FIG. \ref{fig3} and \ref{fig4}, we show the $n=0$ sample solutions for cases of $m=0.025$ and $m=2.0$, respectively. Opposite to vortex solutions, the condensate is non-zero at the origin where the core of the droplet is located, and drops down to zero when approaching to the infinity. So the U(1) gauge symmetry is not broken at large $r$. The magnetic field exhibits new behavior. For high temperatures, it decreases from a non-zero value at the core and down to a constant value at large $r$. For low temperatures, it increases from a non-zero value at the core and all the way to a constant value at large $r$. Similar behavior has also been found for S-wave superconductors in \cite{albash2009a} and \cite{albash2009b}. The magnetic field penetrates almost all of the superconductor except the small region around the core of the droplet. The solutions again behave more like droplets at lower temperatures than at higher temperatures.

Compare vortex solutions with droplet solutions, one can find that droplet solutions form at higher values of the magnetic field. This is consistent to the fact that our D-wave holographic superconductor model is of Type II superconductors. Type II superconductor has two critical magnetic fields, the lower one $B_{c1}$ and the higher one $B_{c2}$. Below $B_{c1}$, it is completely in the superconducting state. Above $B_{c2}$, it is completely in the normal state. In between $B_{c1} <B <B_{c2}$, it is in a mixed state of superconducting and normal states. As $B$ increase from $B_{c1}$, the superconducting state dominates the mixed state, it is preferable for vortex solutions to form first. When $B$ becomes well greater than $B_{c1}$, the normal state dominates the mixed state, then it is expected to find droplet solutions. Of course, it is also possible that the vortex lattice maybe form for some values of the magnetic field. The details remain for future work.

\section{Conclusions}

We have studied the non-trivial localized solutions of a D-wave superconductor model of \cite{benini2010a} and \cite{benini2010b} in the presence of a background magnetic field. Numerically, two types of solutions are found, the vortex and the droplet solutions. The properties of these solutions are discussed. The forming of these solutions depends on the temperature and the magnetic field. They are important configurations in the phase diagram. Of course, a complete investigation of the full phase diagram is missing in this work, which deserves a future study on it. Another interesting project for future study is to try to derive more realistic D-wave holographic models directly from string theory or M-theory, just as what have been tried for S-wave and P-wave superconductors. This method is called the top-down approach. It is interesting to see what new insights of D-wave superconductors the top-down approach may provide us.

\begin{center}
\large{{\bf Acknowledgements}}
\end{center}

This work was supported by fund from Wuhan Institute of Physics and Mathematics, the Chinese Academy of Sciences.

\end{document}